\documentclass[preprint]{aastex631}
\usepackage{xspace}

\def\M87{M87$^*$\xspace}
\def\m87{M87$^*$\xspace}
\def\sgra{Sagittarius~A$^*$\xspace}

\defcitealias{M87PaperI}{Paper~I}
\defcitealias{M87PaperII}{Paper~II}
\defcitealias{M87PaperIII}{Paper~III}
\defcitealias{M87PaperIV}{Paper~IV}
\defcitealias{M87PaperV}{Paper~V}
\defcitealias{M87PaperVI}{Paper~VI}

\shorttitle{New Views of Black Holes from Computational Imaging}
\shortauthors{Akiyama, Chael \& Pesce}

\graphicspath{{./}{figures/}}

\begin{document}

\title{New Views of Black Holes from Computational Imaging}

\correspondingauthor{Kazunori Akiyama}
\author[0000-0002-9475-4254]{Kazunori Akiyama}
\affiliation{Massachusetts Institute of Technology Haystack Observatory, 99 Millstone Road, Westford, MA 01886, USA}
\affiliation{National Astronomical Observatory of Japan, 2-21-1 Osawa, Mitaka, Tokyo 181-8588, Japan}
\affiliation{Black Hole Initiative at Harvard University, 20 Garden Street, Cambridge, MA 02138, USA}
\email{kakiyama@mit.edu}

\author[0000-0003-2966-6220]{Andrew Chael}
\affil{Princeton Center for Theoretical Science, Jadwin Hall, Princeton University, Princeton, NJ 08544, USA}

\author[0000-0002-5278-9221]{Dominic W. Pesce}
\affiliation{Center for Astrophysics $|$ Harvard \& Smithsonian, 60 Garden Street, Cambridge, MA 02138, USA}
\affiliation{Black Hole Initiative at Harvard University, 20 Garden Street, Cambridge, MA 02138, USA}

\section*{ }
\vspace{-2eM}

\begin{center}
	\textit{This is an authors' version of \citet{Akiyama_2021} published from Nature Computational Science.}
\end{center}
\vspace{1em}

\noindent \textbf{The unique challenges associated with imaging a black hole motivated the development of new computational imaging algorithms. As the Event Horizon Telescope continues to expand, these algorithms will need to evolve to keep pace with the increasingly demanding volume and dimensionality of the data.}
\vspace{1em}

\begin{figure}[t]
    \centering
    \includegraphics[width=1.0\textwidth]{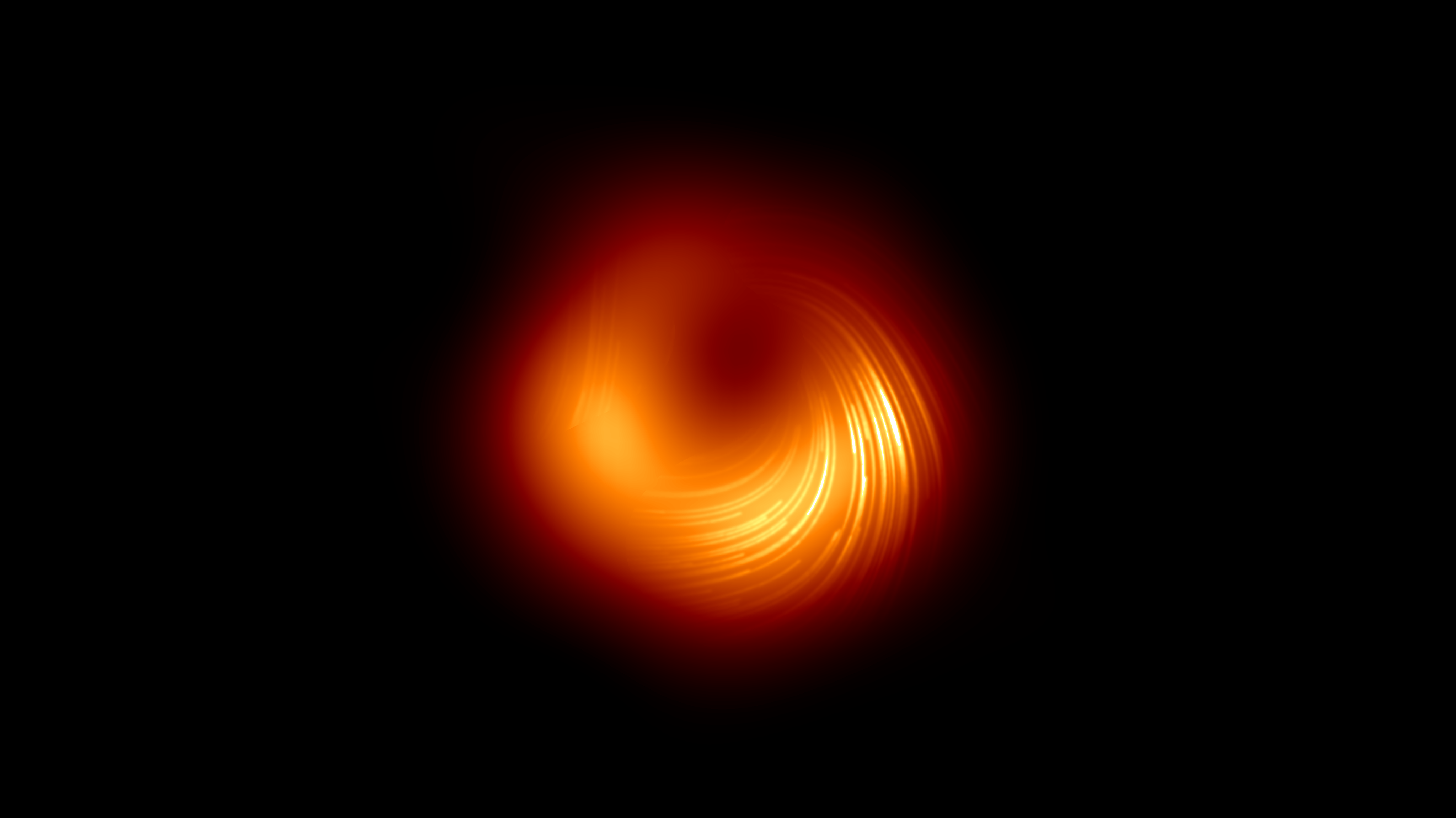}
    \caption{\textbf{The image of the black hole \m87 at the center of the galaxy M87, as revealed by the Event Horizon Telescope collaboration} \citep{M87PaperIV,M87PaperVII}.  The color scale traces the total intensity of the gravitationally lensed radio light emitted from around the black hole, and the streamlines indicate the polarization direction at each point in the image.}
    \label{fig:m87image}
\end{figure}

Two years ago, the Event Horizon Telescope (EHT) collaboration -- an international team of over 200 scientists -- captured the first image of a black hole;
weighing 6.5 billion solar masses, the source of the EHT image, Messier 87$^{*}$ (\m87), is located in the center of the M87 galaxy, 55 million light years away
\citep{M87PaperI,M87PaperII,M87PaperIII,M87PaperIV,M87PaperV,M87PaperVI}. 
Just a few weeks ago, the EHT collaboration revealed an image of that same black hole in polarized light, whose structure is imprinted by the magnetic fields that live at the edge of the black hole horizon \citep{M87PaperVII,M87PaperVIII}.  The images produced by the EHT collaboration show a bright ring of emission encircling a dark central ``shadow'' (\autoref{fig:m87image}), a structure formed as light rays emitted near the black hole
have their trajectories altered (``lensed'') by the intense gravity.
These remarkable scientific breakthroughs were made possible by algorithmic innovations (\autoref{fig:data_flow}) that enabled the data gathered by eight sensitive radio telescopes spread across the globe (\autoref{fig:ehtarray}) to be combined into an image equivalent to what would be seen by a single, Earth-sized telescope. Making the EHT black hole image required addressing computational challenges at every step of the signal pathway, from data collection, correlation, and calibration through imaging and downstream analyses.

\section*{Building up a virtual Earth-sized telescope}
The EHT is currently an array of eleven sensitive radio telescopes (\autoref{fig:ehtarray}); using the technique of Very Long Baseline Interferometry (VLBI) their signals are combined to effectively form a computational planet-wide telescope. 
As cameras adaptively adjust their optics to focus, each part of the ``virtual mirror'' of the giant synthesized telescope needs to be focused against the turbulent atmosphere. 
To do so, the faint radio signals from the black hole source need to be detected in the much louder noise of the telescope receiving systems.
The EHT needed to wait until the late 2010s to outfit a set of sensitive telescopes with the dedicated broadband receiving systems that could capture a sufficient number of radio photons in a short segment of time. 

\begin{figure}[t]
    \centering
    \includegraphics[width=1.0\textwidth]{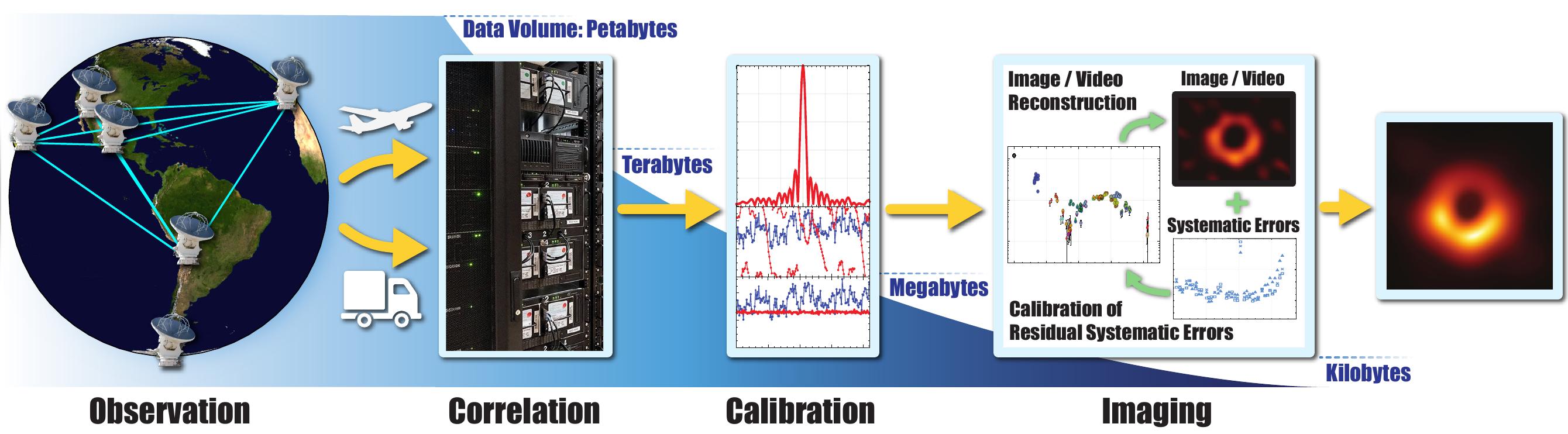}
    \caption{\textbf{A schematic view of the EHT data processing}. The EHT Data processing will recover a black hole image from petabytes of radio signals through the enormous amount of data reduction over twelve order-of-magnitudes down to kilobytes. The computational imaging is the final procedure that reconstructs the black hole image from a sparse set of calibrated measurements together with the remaining systematic calibration errors. Some images are reproduced from \citet{M87PaperII}, \citet{M87PaperIV} and \citet{Blackburn_HOPS_2019}.}
    \label{fig:data_flow}
\end{figure}

\begin{figure}[t]
    \centering
    \includegraphics[width=1.0\textwidth]{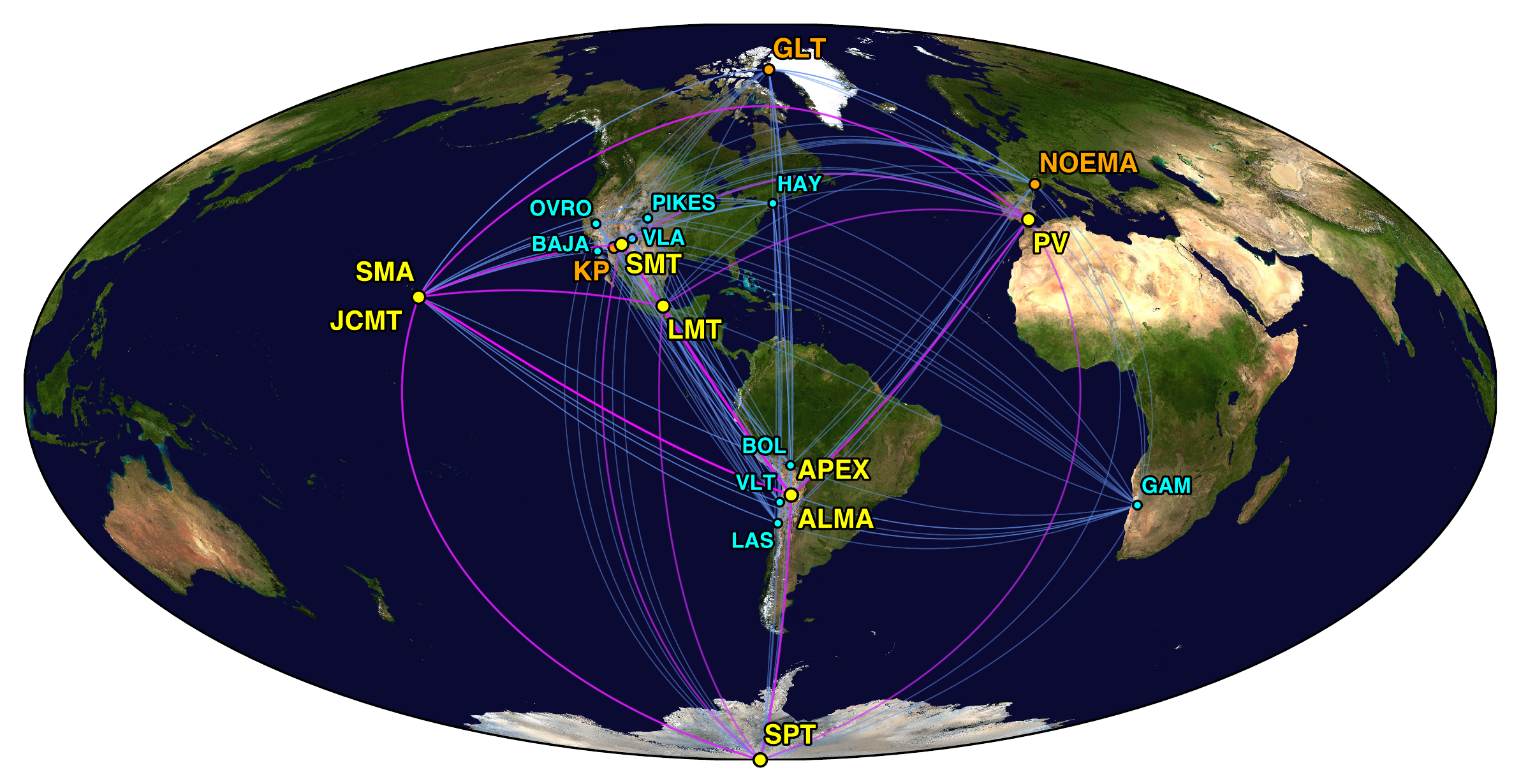}
    \caption{\textbf{The past, current, and future EHT array}.  The first major EHT results were obtained in 2017, using a sparse array of eight telescopes located at six geographic sites \citep{M87PaperII}.  Stations participating in the 2017 configuration are labeled in yellow, and the baselines connecting them are colored magenta.  Since 2017, the EHT has added three additional sites to the array, which are labeled in orange.  Over the coming decade, the next-generation EHT is expect to add a number of additional sites to the array; several possible locations for these new sites are labeled in cyan. The figure is reproduced from \citet[][]{Blackburn_2019}.}
    \label{fig:ehtarray}
\end{figure}

In the EHT, radio waves are collected at a recording rate of several gigabytes per second at each telescope, reaching petabytes of data in total (\autoref{fig:data_flow}). The massive data set, recorded on to hard disk drives, was physically shipped and then processed at supercomputers (called “correlators”)
to extract the signal from the black hole image buried in the telescope system noise; only one part in $10^4$ is the astronomical signal.
The segment of the EHT virtual mirror is formed by pairs of antennas at each time and frequency segment \citep{M87PaperIII}, sampling a Fourier component of the sky image in the ``aperture'' of the effective planet-sized telescope \citep{M87PaperIV}.
The spatial frequency sampled in the Fourier space is determined by the projected separation of the pair of telescopes as seen from the source.
Several new algorithms and computational pipelines were developed to calibrate the data and enhance the sensitivity of the array at this stage \citep{M87PaperIII, Blackburn_HOPS_2019, Janssen_2019}.  
After these calibrations, the data were integrated over time and frequency. This process reduces the data volume from the original petabytes of raw data to only a few megabytes for imaging and scientific analysis.

The radio signals detected by the EHT provide an enormously rich data set that offers the extremely fine resolution necessary to resolve the shadow of a black hole and capture any dynamical evolution of hot plasma around the black hole on timescales of seconds. At the same time, these data are very sparse; they only fill in a  few parts of the Earth-sized virtual aperture.
The inverse problem of reconstructing an image from the measurements is under-constrained, even if the telescope measurements were perfect.
In reality, the imperfect characterization of the sensitivity and turbulent atmospheric scattering of the radio signals at each telescope leave residual 
systematic calibration errors in the data, which need to be solved simultaneously with the unknown image structure (\autoref{fig:data_flow}). 

\section*{The first images of a black hole}
For years leading up to the first observation with the full EHT, the team worked extensively on developing new methods for reconstructing images that deal specifically with the problems of the sparsity  and imperfect calibration of the array. Traditional methods for image reconstruction in radio interferometry,  called CLEAN methods \citep[e.g.][]{Hobgom_1974}, deconvolve artifacts from the array's sparse coverage in the image domain.
To account for telescope calibration errors, these methods iterate between solving for these residual calibration factors and solving for the image \citep{Readhead1978}. 
However, CLEAN methods do not directly incorporate either statistical or residual calibration errors on data into the imaging process. 
In addition to CLEAN methods, the EHT developed new, dedicated approaches to imaging that deal directly with the sparsity and uncertain calibration of the measurements.

A major focus of development was in so-called ``Regularized Maximum Likelihood'' (RML) methods \citep[e.g.][]{Chael_2016,Akiyama_2017a,Akiyama_2017b,Chael_2018}. In this approach, images are fit to the data 
by maximizing a cost term composed of the data likelihood plus ``regularizing'' terms that penalize or favor images with certain features. 
RML methods offer flexibility to the imaging; they can use robust data products that are immune to calibration uncertainties in their likelihood functions. They can also incorporate physically reasonable prior constraints in the regularization terms of the loss function, such as image positivity, sparsity, or smoothness \citep[see][for review]{M87PaperIV}.

Before the first observations of \m87, progress in developing new methods and software for EHT image reconstruction was aided by a spirit of friendly competition  between multiple groups -- in particular, several ``image challenges,'' where teams attempted to obtain the most accurate image reconstruction from a synthetic EHT data set. When the real data was ready, the EHT team decided not to adopt only a single method with producing an image on its own; instead, the team compared multiple techniques, including both traditional tools like CLEAN and new RML algorithms.
To reduce bias and fine-tuning of free parameters in the imaging methods that might push the reconstruction toward more spectacular images, the team broke up into four smaller groups that attempted to reconstruct images from the data independently, with no cross-communication for several weeks.
When the full group met to compare images, they were strikingly consistent -- all featuring a ring of approximately the size predicted from theoretical models of the emission around \m87. 
This first blind validation test, as well as extensive testing of each algorithm's sensitivity to different parameters on synthetic data sets, provided confidence that the final published images were accurate, conservative representations of the EHT data.

\section*{Imaging Magnetic Fields}
The new \emph{polarization} images from the EHT are a major advance from the original images of the total intensity \citep[Figure \ref{fig:m87image};][]{M87PaperVII, M87PaperVIII}. The polarized light from the black hole in M87 originates from relativistic electrons spiraling around magnetic field lines; the direction and magnitude of the polarized signal constrain the strength and structure of the magnetic fields around the black hole. The new information from the polarimetric images considerably narrows the space of allowed models for the structure of the region just outside the black hole's horizon. In particular, the EHT polarization images indicate that the magnetic fields in this region are strong and coherently ordered; they push back against the infalling gas and extract the rotational energy of the black hole to launch a powerful jet outside of the black hole's host galaxy. 

While polarization adds significant new information about the physics of the M87 black hole, it also adds several new dimensions to the imaging problem. 
To make a polarized image from  EHT data requires solving for at least three different images simultaneously instead of one; the total intensity of the radio signal including the unpolarized part, and the magnitude and orientation of the polarization vectors.
In addition, imperfections in the signal chain at the telescopes mix the polarization signals together, introducing additional calibration errors. 

As part of the process of moving to polarized data, the EHT developed a new class of image reconstruction algorithms that attempt to characterize the full space of uncertainties on the polarized image and calibration factors, given uncertainties in the data.
This task is naturally captured by the posterior distribution of images within a Bayesian statistical framework, and realized with new imaging codes based on Markov chain Monte Carlo (MCMC) methods \citep[][]{Broderick_2020a,Pesce_2021} developed within the EHT. While more computationally expensive than gradient-descent based imaging methods, these new tools allow for a full characterization of the uncertainty in the calibration terms and different image structures. 

In producing polarized images, the EHT used these new posterior exploration algorithms together with CLEAN \citep{Marti-Vidal_2021,Park_2021} and RML \citep{Chael_2016} methods.
Following a similar approach as for the first images, extensively testing different computational methods on realistic synthetic data sets and comparing results across methods, the team developed a better understanding of the different strengths and weaknesses of each approach, and built confidence in the final  result as an accurate representation of the magnetic field structure around the black hole.

\section*{New challenges for the next-generation EHT}
The EHT continues to grow, both through the addition of new telescopes to the array and through technological advances that permit 
increased data capture rates at each telescope \citep[\autoref{fig:ehtarray};][]{Blackburn_2019}.  Future EHT observations will also add more dimensions to the data and correspondingly more challenges to the image reconstruction problem.  For instance:
\begin{itemize}
    \item The enhanced array will see emission that is hundreds of times fainter and which extends over a much larger field of view than the bright, compact ring seen in 2017.  Such sensitivity is necessary for capturing the connection between the black hole in M87 and the relativistic jet that it launches, but imaging this faint and extended emission requires algorithms capable of accurately reconstructing images across several orders of magnitude in both brightness and spatial scale. 
    \item High-cadence observations will be required to produce images on short enough timescales to capture the dynamics of \sgra, the black hole in the center of our Milky Way galaxy around which plasma can orbit with periods of only several minutes.  On such short timescales data sparsity is the limiting factor, and dynamic imaging methods must encode the expected temporal correlations to successfully reconstruct ``movies'' from these data \citep[e.g.,][]{Johnson_2017, Bouman_2017, Levis_2021}.
    \item Observations spanning multiple radio frequencies will provide more constraints on the physical conditions in the relativistic plasma around the black hole.
\end{itemize}

In addition to the challenges associated with increased data quality and volume, next-generation EHT imaging algorithms will strive to enable precision tests of general relativity, which require a precise understanding of the uncertainties inherent in any measurements made from EHT data.  Uncertainty in the image structure is driven by a number of factors, including the finite signal-to-noise in the measurements, the sparse filling of the array, and the imperfect data calibration.  These sources of uncertainty are traditionally seen as obstacles that imaging algorithms must overcome during the reconstruction process, and much of the work in the RML imaging methods has focused on developing and implementing a suitable suite of image priors that permit unique image solutions in the face of these issues.

The ability of new MCMC-based \citep[e.g.,][]{Broderick_2020a,Pesce_2021} and variational inference \citep[e.g.,][]{Sun_2020} imaging tools developed within the EHT to analyze a posterior distribution of image structures permits a quantification of the significance and reliability with which various image features are detected.  By grounding the imaging further upstream in the data reduction pipeline, uncertainties associated with calibration choices can be self-consistently incorporated into the posterior distributions on image features \citep[e.g.,][]{Natarajan_2020}.
Critically, by tying the imaging process to downstream science analyses, these posteriors can derive uncertainties on the most important physical quantities in the image \citep[e.g.,][]{Broderick_2020b,Tiede_2020}, potentially enabling more precise tests of astrophysical theories and gravity itself in the near-horizon  environment.

\section*{Outlook}
In the end, the imaging problem is only one of many that the present and future EHT must contend with.
The EHT requires the continued innovation in every single step of the signal processing, from hardware engineering and the human logistical challenges associated with simultaneously operating multiple observatories at hard-to-reach sites around the globe to the computational challenge of calibrating and averaging petabytes of raw data down by orders of magnitude.  From these steps on through imaging and to a myriad of scientific analyses that occur downstream, the ultimate ``imaging algorithm'' would provide a complete end-to-end description of this data trajectory.
For the time being such an algorithm lives only in the idealized realm of our imagination, but we continue to chip away at the edges.

\section*{acknowledgement}
We are sincerely grateful to the editor Fernando Chirigati for the invitation and many suggestions that helped to improve this Comment. We thank K. Bouman, G. Bower, P. Galison, S. Ikeda and M. Janssen for helpful comments. 
K.A. is supported by the National Science Foundation (NSF) through grants AST-1440254, AST-1614868 and AST-2034306. D.W.P is supported by the NSF through grants AST-1952099, AST-1935980, AST-1828513 and AST-1440254, and by the Gordon and Betty Moore Foundation through grant GBMF5278. 
A.C. is supported by Hubble Fellowship grant HST-HF2-51431.001-A awarded by the Space Telescope Science Institute, which is operated by the Association of Universities for Research in Astronomy, Inc., for NASA, under contract NAS5-26555. 
The Black Hole Initiative at Harvard University is funded by grants from the John Templeton Foundation and the Gordon and Betty Moore Foundation to Harvard University.

\section*{author contributions}
The three authors equally contributed to the Comment. 
Each third of the Comment was drafted by each author, followed by thorough edits by all three authors. K.A. created Figs. 1 and 2; D.W.P created Fig. 3.


\begin{thebibliography}{}
	\expandafter\ifx\csname natexlab\endcsname\relax\def\natexlab#1{#1}\fi
	\providecommand{\url}[1]{\href{#1}{#1}}
	\providecommand{\dodoi}[1]{doi:~\href{http://doi.org/#1}{\nolinkurl{#1}}}
	\providecommand{\doeprint}[1]{\href{http://ascl.net/#1}{\nolinkurl{http://ascl.net/#1}}}
	\providecommand{\doarXiv}[1]{\href{https://arxiv.org/abs/#1}{\nolinkurl{https://arxiv.org/abs/#1}}}
	
	\bibitem[{{Akiyama} {et~al.}(2021){Akiyama}, {Chael}, \&
		{Pesce}}]{Akiyama_2021}
	{Akiyama}, K., {Chael}, A., \& {Pesce}, D.~W. 2021, Nature Computational
	Science, \dodoi{10.1038/s43588-021-00078-z}
	
	\bibitem[{{Akiyama} {et~al.}(2017{\natexlab{a}}){Akiyama}, {Kuramochi},
		{Ikeda}, {Fish}, {Tazaki}, {Honma}, {Doeleman}, {Broderick}, {Dexter},
		{Mo{\'s}cibrodzka}, {Bouman}, {Chael}, \& {Zaizen}}]{Akiyama_2017a}
	{Akiyama}, K., {Kuramochi}, K., {Ikeda}, S., {et~al.} 2017{\natexlab{a}}, \apj,
	838, 1, \dodoi{10.3847/1538-4357/aa6305}
	
	\bibitem[{{Akiyama} {et~al.}(2017{\natexlab{b}}){Akiyama}, {Ikeda}, {Pleau},
		{Fish}, {Tazaki}, {Kuramochi}, {Broderick}, {Dexter}, {Mo{\'s}cibrodzka},
		{Gowanlock}, {Honma}, \& {Doeleman}}]{Akiyama_2017b}
	{Akiyama}, K., {Ikeda}, S., {Pleau}, M., {et~al.} 2017{\natexlab{b}}, \aj, 153,
	159, \dodoi{10.3847/1538-3881/aa6302}
	
	\bibitem[{{Blackburn} {et~al.}(2019){Blackburn}, {Chan}, {Crew}, {Fish},
		{Issaoun}, {Johnson}, {Wielgus}, {Akiyama}, {Barrett}, {Bouman}, {Cappallo},
		{Chael}, {Janssen}, {Lonsdale}, \& {Doeleman}}]{Blackburn_HOPS_2019}
	{Blackburn}, L., {Chan}, C.-k., {Crew}, G.~B., {et~al.} 2019, \apj, 882, 23,
	\dodoi{10.3847/1538-4357/ab328d}
	
	\bibitem[{Bouman {et~al.}(2018)Bouman, Johnson, Dalca, Chael, Roelofs,
		Doeleman, \& Freeman}]{Bouman_2017}
	Bouman, K.~L., Johnson, M.~D., Dalca, A.~V., {et~al.} 2018, IEEE Transactions
	on Computational Imaging, 4, 512, \dodoi{10.1109/TCI.2018.2838452}
	
	\bibitem[{{Broderick} {et~al.}(2020{\natexlab{a}}){Broderick}, {Gold},
		{Karami}, {Preciado-L{\'o}pez}, {Tiede}, {Pu}, {Akiyama}, \&
		et~al.}]{Broderick_2020a}
	{Broderick}, A.~E., {Gold}, R., {Karami}, M., {et~al.} 2020{\natexlab{a}},
	\apj, 897, 139, \dodoi{10.3847/1538-4357/ab91a4}
	
	\bibitem[{{Broderick} {et~al.}(2020{\natexlab{b}}){Broderick}, {Pesce},
		{Tiede}, {Pu}, \& {Gold}}]{Broderick_2020b}
	{Broderick}, A.~E., {Pesce}, D.~W., {Tiede}, P., {Pu}, H.-Y., \& {Gold}, R.
	2020{\natexlab{b}}, \apj, 898, 9, \dodoi{10.3847/1538-4357/ab9c1f}
	
	\bibitem[{{Chael} {et~al.}(2018){Chael}, {Johnson}, {Bouman}, {Blackburn},
		{Akiyama}, \& {Narayan}}]{Chael_2018}
	{Chael}, A.~A., {Johnson}, M.~D., {Bouman}, K.~L., {et~al.} 2018, \apj, 857,
	23, \dodoi{10.3847/1538-4357/aab6a8}
	
	\bibitem[{{Chael} {et~al.}(2016){Chael}, {Johnson}, {Narayan}, {Doeleman},
		{Wardle}, \& {Bouman}}]{Chael_2016}
	{Chael}, A.~A., {Johnson}, M.~D., {Narayan}, R., {et~al.} 2016, \apj, 829, 11,
	\dodoi{10.3847/0004-637X/829/1/11}
	
	\bibitem[{{Doeleman} {et~al.}(2019){Doeleman}, {Blackburn}, {Dexter}, {Gomez},
		{Johnson}, {Palumbo}, {Weintroub}, {Farah}, {Fish}, {Loinard}, {Lonsdale},
		{Narayanan}, {Patel}, {Pesce}, {Raymond}, \& et~al.}]{Blackburn_2019}
	{Doeleman}, S., {Blackburn}, L., {Dexter}, J., {et~al.} 2019, in Bulletin of
	the American Astronomical Society, Vol.~51, 256.
	\newblock \doarXiv{1909.01411}
	
	\bibitem[{{Event Horizon Telescope Collaboration}
		{et~al.}(2019{\natexlab{a}}){Event Horizon Telescope Collaboration},
		{Akiyama}, {Alberdi}, {Alef}, {Asada}, {Azulay}, {Baczko}, {Ball},
		{Balokovi{\'c}}, {Barrett}, \& et~al.}]{M87PaperI}
	{Event Horizon Telescope Collaboration}, {Akiyama}, K., {Alberdi}, A., {et~al.}
	2019{\natexlab{a}}, \apjl, 875, L1, \dodoi{10.3847/2041-8213/ab0ec7}
	
	\bibitem[{{Event Horizon Telescope Collaboration}
		{et~al.}(2019{\natexlab{b}}){Event Horizon Telescope Collaboration},
		{Akiyama}, {Alberdi}, {Alef}, {Asada}, {Azulay}, {Baczko}, {Ball},
		{Balokovi{\'c}}, {Barrett}, \& et~al.}]{M87PaperII}
	---. 2019{\natexlab{b}}, \apjl, 875, L2, \dodoi{10.3847/2041-8213/ab0c96}
	
	\bibitem[{{Event Horizon Telescope Collaboration}
		{et~al.}(2019{\natexlab{c}}){Event Horizon Telescope Collaboration},
		{Akiyama}, {Alberdi}, {Alef}, {Asada}, {Azulay}, {Baczko}, {Ball},
		{Balokovi{\'c}}, {Barrett}, \& et~al.}]{M87PaperIII}
	---. 2019{\natexlab{c}}, \apjl, 875, L3, \dodoi{10.3847/2041-8213/ab0c57}
	
	\bibitem[{{Event Horizon Telescope Collaboration}
		{et~al.}(2019{\natexlab{d}}){Event Horizon Telescope Collaboration},
		{Akiyama}, {Alberdi}, {Alef}, {Asada}, {Azulay}, {Baczko}, {Ball},
		{Balokovi{\'c}}, {Barrett}, \& et~al.}]{M87PaperIV}
	---. 2019{\natexlab{d}}, \apjl, 875, L4, \dodoi{10.3847/2041-8213/ab0e85}
	
	\bibitem[{{Event Horizon Telescope Collaboration}
		{et~al.}(2019{\natexlab{e}}){Event Horizon Telescope Collaboration},
		{Akiyama}, {Alberdi}, {Alef}, {Asada}, {Azulay}, {Baczko}, {Ball},
		{Balokovi{\'c}}, {Barrett}, \& et~al.}]{M87PaperV}
	---. 2019{\natexlab{e}}, \apjl, 875, L5, \dodoi{10.3847/2041-8213/ab0f43}
	
	\bibitem[{{Event Horizon Telescope Collaboration}
		{et~al.}(2019{\natexlab{f}}){Event Horizon Telescope Collaboration},
		{Akiyama}, {Alberdi}, {Alef}, {Asada}, {Azulay}, {Baczko}, {Ball},
		{Balokovi{\'c}}, {Barrett}, \& et~al.}]{M87PaperVI}
	---. 2019{\natexlab{f}}, \apjl, 875, L6, \dodoi{10.3847/2041-8213/ab1141}
	
	\bibitem[{{Event Horizon Telescope Collaboration}
		{et~al.}(2021{\natexlab{a}}){Event Horizon Telescope Collaboration},
		{Akiyama}, {Alberdi}, {Alef}, {Asada}, {Azulay}, {Baczko}, {Ball},
		{Balokovi{\'c}}, {Barrett}, \& et~al.}]{M87PaperVII}
	---. 2021{\natexlab{a}}, \apjl, 910, L12, \dodoi{10.3847/2041-8213/abe71d}
	
	\bibitem[{{Event Horizon Telescope Collaboration}
		{et~al.}(2021{\natexlab{b}}){Event Horizon Telescope Collaboration},
		{Akiyama}, {Alberdi}, {Alef}, {Asada}, {Azulay}, {Baczko}, {Ball},
		{Balokovi{\'c}}, {Barrett}, \& et~al.}]{M87PaperVIII}
	---. 2021{\natexlab{b}}, \apjl, 910, L13, \dodoi{10.3847/2041-8213/abe4de}
	
	\bibitem[{{H{\"o}gbom}(1974)}]{Hobgom_1974}
	{H{\"o}gbom}, J.~A. 1974, \aaps, 15, 417
	
	\bibitem[{{Janssen} {et~al.}(2019){Janssen}, {Goddi}, {van Bemmel}, {Kettenis},
		{Small}, {Liuzzo}, {Rygl}, {Mart{\'\i}-Vidal}, {Blackburn}, {Wielgus}, \&
		{Falcke}}]{Janssen_2019}
	{Janssen}, M., {Goddi}, C., {van Bemmel}, I.~M., {et~al.} 2019, \aap, 626, A75,
	\dodoi{10.1051/0004-6361/201935181}
	
	\bibitem[{{Johnson} {et~al.}(2017){Johnson}, {Bouman}, {Blackburn}, {Chael},
		{Rosen}, {Shiokawa}, {Roelofs}, {Akiyama}, {Fish}, \&
		{Doeleman}}]{Johnson_2017}
	{Johnson}, M.~D., {Bouman}, K.~L., {Blackburn}, L., {et~al.} 2017, \apj, 850,
	172, \dodoi{10.3847/1538-4357/aa97dd}
	
	\bibitem[{{Levis} {et~al.}(2021){Levis}, {Lee}, {Tropp}, {Gammie}, \&
		{Bouman}}]{Levis_2021}
	{Levis}, A., {Lee}, D., {Tropp}, J.~A., {Gammie}, C.~F., \& {Bouman}, K.~L.
	2021, submitted to ICCV 2021
	
	\bibitem[{{Mart{\'\i}-Vidal} {et~al.}(2021){Mart{\'\i}-Vidal}, {Mus},
		{Janssen}, {de Vicente}, \& {Gonz{\'a}lez}}]{Marti-Vidal_2021}
	{Mart{\'\i}-Vidal}, I., {Mus}, A., {Janssen}, M., {de Vicente}, P., \&
	{Gonz{\'a}lez}, J. 2021, \aap, 646, A52, \dodoi{10.1051/0004-6361/202039527}
	
	\bibitem[{{Natarajan} {et~al.}(2020){Natarajan}, {Deane}, {van Bemmel}, {van
			Langevelde}, {Small}, {Kettenis}, {Paragi}, {Smirnov}, \&
		{Szomoru}}]{Natarajan_2020}
	{Natarajan}, I., {Deane}, R., {van Bemmel}, I., {et~al.} 2020, \mnras, 496,
	801, \dodoi{10.1093/mnras/staa1503}
	
	\bibitem[{{Park} {et~al.}(2021){Park}, {Byun}, {Asada}, \& {Yun}}]{Park_2021}
	{Park}, J., {Byun}, D.-Y., {Asada}, K., \& {Yun}, Y. 2021, \apj, 906, 85,
	\dodoi{10.3847/1538-4357/abcc6e}
	
	\bibitem[{{Pesce}(2021)}]{Pesce_2021}
	{Pesce}, D.~W. 2021, \aj, 161, 178, \dodoi{10.3847/1538-3881/abe3f8}
	
	\bibitem[{{Readhead} \& {Wilkinson}(1978)}]{Readhead1978}
	{Readhead}, A.~C.~S., \& {Wilkinson}, P.~N. 1978, \apj, 223, 25,
	\dodoi{10.1086/156232}
	
	\bibitem[{Sun {et~al.}(2020)Sun, Dalca, \& Bouman}]{Sun_2020}
	Sun, H., Dalca, A.~V., \& Bouman, K.~L. 2020, in 2020 IEEE International
	Conference on Computational Photography (ICCP), 1--12,
	\dodoi{10.1109/ICCP48838.2020.9105133}
	
	\bibitem[{{Tiede} {et~al.}(2020){Tiede}, {Broderick}, \&
		{Palumbo}}]{Tiede_2020}
	{Tiede}, P., {Broderick}, A.~E., \& {Palumbo}, D. C.~M. 2020, arXiv e-prints,
	arXiv:2012.07889.
	\newblock \doarXiv{2012.07889}
	
\end{thebibliography}

\end{document}